\documentstyle[11pt]{article}

\newcommand{\nc}{\newcommand}
\newcommand{\rnc}{\renewcommand}
\def\a{& \hspace{-7pt}}
\def\Z{{\bf Z}}
\nc{\be}{\begin{equation}}
\nc{\ee}{\end{equation}}
\nc{\bea}{\begin{eqnarray}}
\nc{\eea}{\end{eqnarray}}
\def\nn{\nonumber}
\def\Dslash{\raisebox{1pt}{$\slash$} \hspace{-8pt} D}
\def\Dslashs{\raisebox{0.75pt}{$\scriptstyle{\slash}$} \hspace{-6pt} D}
\def\0{\scriptscriptstyle{(}\raisebox{-1pt}{$\scriptstyle{0}$}
\scriptscriptstyle{)}}
\def\1{\scriptscriptstyle{(}\raisebox{-1pt}{$\scriptstyle{1}$}
\scriptscriptstyle{)}}
\nc{\tr}{\mbox{\begin{rmfamily}tr\end{rmfamily}}}
\nc{\Tr}{\mbox{\begin{rmfamily}Tr\end{rmfamily}}}
\nc{\ch}{\mbox{\begin{rmfamily}ch\end{rmfamily}}}
\nc{\Td}{\mbox{\begin{rmfamily}Td\end{rmfamily}}}
\rnc{\index}{\mbox{\begin{rmfamily}index\end{rmfamily}}}

\begin{document}

\begin{center}

{$\quad$}

\vskip 20pt

{\LARGE \bf Anomaly inflow and RR anomalous \\[2mm] couplings}

\vskip 30pt

{\large Claudio A. Scrucca$^{a}$ and Marco Serone$^{b,c}$}

\vskip 15pt

${}^a$
{\em Sektion Physik, Ludwig Maximilian Universit\"at M\"unchen}\\
{\em Theresienstra\ss e 37, 80333 Munich, Germany}\\

\vskip 2pt

${}^b$
{\em Department of Mathematics, University of Amsterdam}\\
{\em Plantage Muidergracht 24, 1018 TV Amsterdam, The Netherlands} \\

\vskip 2pt

${}^c$
{\em Spinoza Institute, University of Utrecht} \\
{\em Leuvenlaan 4, 3584 CE Utrecht, The Netherlands}

\end{center}

\vskip 10pt

\centerline{\bf Abstract} \vskip 5pt

\begin{small}

We review the anomaly inflow mechanism on D-branes and 
O-planes. In particular, we compute the one-loop world-volume 
anomalies and derive the RR anomalous couplings required for 
their cancellation. 

\end{small}

\vskip 10pt

\section{Anomalies and inflow}

It is known that a consistent quantum field theory can happen to admit as
vacuum a topological defect carrying chiral zero modes. The anomaly arising 
on the world-volume must then be canceled by an inflow from the bulk \cite{ch}.
This is the case of consistent superstring vacua with D-branes and O-planes, 
where no overall anomaly can arise but zero modes occur. In general, there 
can be a net world-volume quantum anomaly, but by consistency, this must be 
canceled by an equal and opposite classical inflow of anomaly.

The W-Z consistency condition implies that any anomaly ${\cal A}$ is the 
descent of some polynomial $I(F,R)$ of the curvatures $F$ and $R$ of the 
gauge and the tangent bundles. Defining $I = dI^{\0}$ and 
$\delta I^{\0} = dI^{\1}$: ${\cal A} = 2 \pi i \int I^{\1}$.
$I(F,R)$ depends on characteristic classes, like ($\lambda_a$ are the 
skew-eigenvalues of $R$)
\bea
\a\a \ch (F) = \tr \, \exp i F / 2 \pi \;, \nn \\
\a\a \widehat A (R) = \prod_{a=1}^{D/2} 
\frac {\lambda_a/4 \pi}{\sinh \lambda_a/4 \pi} 
\;,\;\; \widehat L(R) = \prod_{a=1}^{D/2} 
\frac {\lambda_a / 2 \pi}{\tanh \lambda_a / 2 \pi} 
\;,\;\; e(R) = \prod_{a=1}^{D/2} \lambda_a / 2 \pi \;. \nn
\eea

Beside quantum anomalies, arising from fluctuations of chiral 
fermions or self-dual tensor fields, also classical anomalies 
can occur, for instance in magnetic interactions. Consider indeed
some defects $M_i$ in spacetime $X$, with the RR couplings:
$$
S = - \sum\nolimits_i \mu_i \int_{M_i} C \wedge Y_i \;,
$$
where $C=\sum_p C_{\scriptscriptstyle{(}\raisebox{-1pt}{$\scriptstyle{p}$}
\scriptscriptstyle{)}}$ and $Y=Y(F,R)$.
This can be written as an integral over $X$ by using the 
currents $\tau_{M_i}$, which are globally determined by $N(M_i)$
and locally given by $\tau_{M_i} \sim \delta(x^{d_i})\,dx^{d_i} \wedge ... 
\wedge \delta(x^D)\,dx^D$ \cite{cy}. 
The RR equation of motion and Bianchi identity become then
(the bar represents complex conjugation)
\bea
\a\a d^*\!H = \sum\nolimits_i \mu_i \, \tau_{M_i} \wedge Y_i \;, \nn \\
\a\a d H = - \sum\nolimits_i \mu_i \, \tau_{M_i} \wedge \bar  Y_i \;. \nn
\eea
The modified Bianchi identity implies that 
$H = dC - \sum_i \mu_i \, \tau_{M_i} \wedge \bar  Y_i^{\0}$, 
and since this must be gauge invariant, $C$ must transform
as $\delta C = \sum_i \mu_i \, \tau_{M_i} \wedge \bar  Y_i^{\1}$.
Consequently, the RR couplings give the anomaly
$$
{\cal A} = - \,i\sum\nolimits_{i,j} \mu_i \, \mu_j \int_X \tau_{M_i} 
\wedge \tau_{M_j} \wedge \left(Y_i \wedge \bar  Y_j \right)^{(1)} \;.
$$
Using the property 
$\tau_{M_i} \wedge \tau_{M_j} = \tau_{M_{ij}} \wedge e[N(M_{ij})]$  
\cite{cy}, we see that the magnetic interaction between $M_i$ and $M_j$ 
is anomalous on the intersection $M_{ij}$. The classical anomaly inflow 
on each intersection is
${\cal A}_{ij} = 2 \pi i \int_{M_{ij}} I^{\1}_{ij}$, with
\be
I_{ij} = - \frac {\mu_i \, \mu_j}{2 \pi} \, Y_i \wedge \bar  Y_j 
\wedge e[N(M_{ij})] \;.
\label{inf}
\ee
This must cancel the corresponding quantum anomaly \cite{ghm,cy} 
(even if, strictly speaking, subtletes could arise for self-dual 
objects \cite{cy}).

\section{Anomalies on D-branes and O-planes}

Consider two parallel Dp-branes (B) and/or Op-planes (B) on $M$. 
The anomalous fields living on their world-volumes can be read from 
the corresponding potentially divergent one-loop amplitudes: the 
annulus, the M\"obius strip and the Klein bottle for the BB, BO and OO 
configurations.
In the first two cases, one finds chiral $R$ spinors, and in the 
last case self-dual RR forms. These fields are dimensionally reduced 
from $D=10$ to $D=p+1$, and will therefore split into two sets with
opposite chirality or self-duality. Anomalies can then arise only when 
$N(M)$ is non-trivial. 

These anomalies are as usual topological indices, which can be computed 
using index theorems or via a path-integral representation as in \cite{agw}. 
In this second approach, the topological nature of the results
is related to supersymmetry, and the tangent, normal and gauge bundle 
curvatures are realized in terms of fermionic zero modes as
($M,N,... \in X$; $\mu,\nu,... \in M$; $i,j,... \in N$)
\bea
\a\a R_{\mu \nu} = \frac 12 \hspace{1pt} R_{\mu \nu \rho \sigma} (x_0) 
\hspace{1pt} \psi_0^{\rho} \psi_0^{\sigma} \;,\;\;
R^\prime_{i j} = \frac 12 \hspace{1pt} R_{i j \rho \sigma} (x_0) 
\hspace{1pt} \psi_0^{\rho} \psi_0^{\sigma} \;, \nn \\
\a\a F = \frac 12 \hspace{1pt} F_{\mu \nu} (x_0) 
\hspace{1pt} \psi_0^{\mu} \psi_0^{\nu} \;.
\label{curv}
\eea

\subsection{Chiral spinors}

The anomaly of a chiral spinor reduced from $X$ to $M$ is
$$
{\cal A} = \lim_{t \rightarrow 0} \Tr \,\big[\hspace{1pt}\Gamma^{D+1}
\,\delta\,e^{\raisebox{1pt}{$\scriptstyle{- t (i \Dslashs)^2}$}}\hspace{1pt}
\big] \;.
$$
The trace is over the eigenstates of $i \Dslash$ on $M$, $\Gamma^{D+1}$ is 
the chiral matrix on $X$, and $\delta$ is the operator representing gauge 
transformations. By exponentiating $\delta$, as in \cite{agw}, this can 
be written as ${\cal A} = 2 \pi i Z^{\1}$, where
$$
Z =  \lim_{t \rightarrow 0} \Tr \,\big[\hspace{1pt}\Gamma^{D+1}\,
e^{\raisebox{1pt}{$\scriptstyle{- t (i \Dslashs)^2}$}}\hspace{1pt}\big] \;.
$$

Mathematically, $Z$ is the index of a twisted spin complex:
$Z = \index(i \Dslash)$. The original chiral or anti-chiral spinor 
on $X$ is a section of $S^\pm_{T(X)}$. On $M \subset X$, the
tangent bundle decomposes into tangent and normal components and 
these spin bundles reduce to 
$E^\pm = \big(S^\pm_{T(M)} \otimes S^+_{N(M)}\big) 
\oplus \big(S^\mp_{T(M)} \otimes S^-_{N(M)}\big)$.
Considering also an extra gauge bundle $V$, one has then
the two-term complex
$$
i \Dslash :\; \Gamma \big[M, E^+ \otimes V\big] \rightarrow 
\Gamma \big[M, E^- \otimes V\big] \;.
$$
The index theorem for this spin complex reads 
$$
\index(i \Dslash) = \int_M \ch(V) \, \big[\ch (E^+) - \ch (E^-) \big]\,
\frac {\Td[\hspace{1pt}T(M^{C})\hspace{1pt}]}
{e[\hspace{1pt}T(M)\hspace{1pt}]} \;,
$$
and explicit evaluation yields \cite{cy}
\be
Z = \int_M  \ch(F) \wedge \frac{\widehat{A}(R)}{\widehat{A}(R^{\prime})} 
\wedge e(R^{\prime}) \;.
\label{s}
\ee

Physically, $Z$ can be interpreted as a partition function. More precisely,
for a super quantum mechanics (SQM) with $Q = i \Dslash$ and 
$(-1)^F = \Gamma^{D+1}$, $Z$ would becomes the Witten index \cite{w}:
$$
Z = \Tr\, \big[\hspace{1pt}(-1)^F\,
e^{\raisebox{1pt}{$\scriptstyle{-t H}$}}\hspace{1pt}\big] \;.
$$
The appropriate SQM is obtained by dimensionally reducing the supersymmetric
non-linear sigma model (SNSM) form $D=1+1$ to $D=0+1$ with Neumann and 
Dirichlet boundary conditions $\parallel$ and $\perp$ to $M$:
$x^i=0$, $\psi_1^{\mu} = \psi_2^{\mu}$, $\psi_1^{i} = -\psi_2^{i}$. 
The Lagrangian is:
\bea
L \a=\a \frac 12 \hspace{1pt} g_{\mu\nu}\,\dot x^\mu \dot x^\nu
+ \frac i2 \hspace{1pt} \psi_{\mu} \Big(\dot \psi^{\mu} 
+ \omega_{\rho \;\;\, \nu}^{\;\; \mu} 
\,\dot x^\rho\, \psi^{\nu} \Big)
+ \hspace{1pt} \frac i2 \hspace{1pt} \psi_{i} 
\Big(\dot \psi^{i} + \omega_{\rho \;\; j}^{\;\; i}\, 
\dot x^\rho\, \psi^{j} \Big) \nn \\
\a\;\a + \frac 14 \hspace{1pt} R_{\mu \nu i j} 
\hspace{1pt} \psi^{\mu} \psi^{\nu} \psi^{i} \psi^{j}
+ ... \nn
\eea
where the dots stand for standard terms accounting for the gauge background.
Due to $(-1)^F$, all the fields are periodic and
$$
Z = \int_P \! {\cal D}x^\mu \!
\int_P \! {\cal D} \psi^{\mu} \! 
\int_P {\cal D} \! \psi^{i} \; e^{\raisebox{1pt}
{$\scriptstyle{- S(t)}$}} \;.
$$
For $t \rightarrow 0$, $Z$ is dominated by constant paths with only small
fluctuations: $x^\mu = x_0^\mu + \xi^{\mu}$, $\psi^{\mu} = 
\psi_0^{\mu} + \lambda^{\mu}$, $\psi^{i} = \psi_0^{i} + \lambda^{i}$.
It is enough to keep quadratic interactions and only terms with
the maximum number of $\psi_0$'s, and one finds
$$
L^{eff} = \frac 12 \Big[\dot \xi_{\mu} \dot \xi^{\mu} 
+ i \lambda_{\mu} \dot \lambda^{\mu} 
+ i \lambda_{i} \dot \lambda^{i}
+ i R_{\mu \nu} \hspace{1pt} \dot \xi^{\mu} \xi^{\nu} 
+ R^\prime_{i j} \hspace{1pt} \lambda^{i} \lambda^{j} \Big]
+ \frac 12 \hspace{1pt} R^\prime_{i j} \hspace{1pt}
\psi_0^{i} \psi_0^{j} + i F \;,
$$
with $R$, $R^\prime$ and $F$ given by (\ref{curv}).
The path-integral gives then
\bea
Z \a=\a \int \! d x_0^\mu \! \int \! d\psi_0^{\mu} 
\, \tr \exp \left\{i F t \right\} \, \frac 
{\det\nolimits_P(i \eta_{\mu \nu} \partial_\tau)}
{\det\nolimits_P(\eta_{\mu \nu} \partial_\tau^2 +
i R_{\mu \nu} \partial_\tau)} \nn \\ \a\;\a
\det\nolimits_P(i \eta_{i j} \partial_\tau  
+ R^\prime_{i j}) \int \! d\psi_0^{i} 
\, \exp \left\{\frac t2 R^\prime_{i j} 
\psi_0^{i} \psi_0^{j}\right\} \;. \nn
\eea
Evaluating the determinants, one recovers finally (\ref{s}) \cite{ss1}.

\subsection{Self-dual tensors}

The anomaly of a self-dual tensor reduced from $X$ to $M$ can be written as
$$
A = \frac 14\,\lim_{t \rightarrow 0} \,\Tr 
\,\big[\hspace{1pt}I *_D\,\delta\, 
e^{\raisebox{1pt}{$\scriptstyle{- t \hspace{1pt} {\cal D}^2}$}}
\hspace{1pt}\big] \;,
$$
where $*_D$ is the Hodge operator on $X$ and the trace is over the 
eigenstates of ${\cal D} = d + d^\dagger$. The dynamics is constrained to 
$M \subset X$ thanks to the transverse reflection $I$.
As before, this can be written as
${\cal A} = 2 \pi i Z^{(1)}$, with
$$
Z = - \frac 18 \,\lim_{t \rightarrow 0} \,\Tr 
\,\big[\hspace{1pt}I *_D e^{\raisebox{1pt}{$\scriptstyle{- t
\hspace{1pt}{\cal D}^2}$}}\hspace{1pt}\big] \;.
$$

Mathematically, $Z$ is in this case a $G$-index of the usual
signature complex. Indeed, $Z = - 1/8 \, \index({\cal D}_+^G)$, where
\bea
\a\a {\cal D}_+ :\;  \Gamma \big[X,{}^+\!\!\wedge T^*X \big] 
\longrightarrow \Gamma \big[X,{}^-\!\!\wedge T^*X\big] \;, \nn \\
\a\a G: X \longrightarrow X \; 
\big(I : (x^\mu,x^i) \longrightarrow (x^\mu,-x^i)\big) \;. \nn
\eea
Notice that $G=\Z_2$ is orientation-preserving, as it should, since 
$D$ and $d$ must be even. It leaves $M \subset X$ fixed and acts as 
$+{\bf 1}$ in $T(M)$ and $-{\bf 1}$ in $N(M)$.
The $G$-signature theorem gives then
$$
\index({\cal D}_+^G)= \int_{M}\, \frac{\big[\ch (E^+) - \ch (E^-) \big]\,
\big[\ch (F^+) - \ch (F^-) \big]}{\ch (F)} \, 
\frac {\Td[\hspace{1pt}T(M^{C})\hspace{1pt}]}
{e[\hspace{1pt}T(M)\hspace{1pt}]} \;,
$$
with the definitions $E^\pm = {}^\pm\!\!\wedge T^*M$, 
$F^\pm = {}^\pm\!\!\wedge N^*M$ and $F = \oplus_i (-1)^i \wedge^i N^*M$. 
One finds finally \cite{ss1}
\be
Z = - \frac 18 
\int_M \frac{\widehat{L}(R)}{\widehat{L}(R^{\prime})} \wedge e(R^{\prime}) \;.
\label{t} 
\ee

Physically, $Z$ looks again like a partition function, and for a SQM with 
$H = {\cal D}^2$ and a symmetry $\Omega = *_D$, it would becomes the 
supersymmetric index
$$
Z = - \frac 18 \Tr\,\big[\hspace{1pt}I\,\Omega\,
e^{\raisebox{1pt}{$\scriptstyle{- t H}$}}
\hspace{1pt}\big] \;.
$$
The appropriate SQM is known \cite{w} to be the trivial dimensional reduction 
of the SNSM from $D=1+1$ to $D=0+1$ ($\Omega: (\psi_1, \psi_2) 
\rightarrow (-\psi_1, \psi_2)$):
\bea
L \a=\a \frac 12 \hspace{1pt} g_{MN} \hspace{1pt} (x)
\hspace{1pt} \dot{x}^M\dot{x}^N \!+ 
\frac i2 \! \sum_{\alpha=1,2} \psi_{\alpha M} 
\Big(\dot \psi_{\alpha}^{M} \!+
\omega_{P\;\;\,N}^{\;\;\,M}\hspace{1pt}(x)\hspace{1pt}
\psi_{\alpha}^{N}\,\dot{x}^P \Big) \nn \\ \a\;\a
+ \frac 14 R_{M N P Q} \hspace{1pt}(x)\hspace{1pt}
\psi_1^{M}\psi_1^{N}\psi_2^{P}\psi_2^{Q}
\;. \nn
\eea
Due to $I\,\Omega$, the fields acquire non-standard periodicities and
$$
Z = - \frac 18 \int_P \! {\cal D}x^\mu \!
\int_A \! {\cal D}x^{i} \!
\int_P \! {\cal D} \psi_1^{\mu} \! 
\int_A \! {\cal D} \psi_1^{i} \!
\int_A {\cal D} \! \psi_2^{\mu} \!
\int_P {\cal D} \! \psi_2^{i} \; e^{\raisebox{1pt}
{$\scriptstyle{- S(t)}$}} \;.
$$
For $t \rightarrow 0$, $Z$ is again dominated by constant paths with small 
fluctuations: $x^\mu = x_0^\mu + \xi^{\mu}$, $x^i = \xi^i$, 
$\psi_1^{\mu} = \psi_{0}^{\mu } + \lambda_1^{\mu}$, 
$\psi_1^{i} = \lambda_1^{i}$, $\psi_2^{\mu} = \lambda_2^{\mu}$,
$\psi_2^{i} = \psi_{0}^{i} + \lambda_2^{i}$.
As before, it is enough to keep terms quadratic in the fluctuations and 
with a maximum number of fermionic zero modes. One finds 
(with $R$, $R^\prime$ as in (\ref{curv}))
\bea
L^{eff} \a=\a 
\frac 12 \Big[\hspace{1pt}\dot \xi_{\mu} \dot \xi^{\mu} 
+ \dot \xi_{i} \dot \xi^{i}
+ i \lambda_{1 \mu} \dot \lambda_1^{\mu}  
+ i \lambda_{1 i} \dot \lambda_1^{i}
+ i \lambda_{2 \mu} \dot \lambda_2^{\mu}  
+ i \lambda_{2 i} \dot \lambda_2^{i} \nn \\ 
\a\;\a \hspace{17pt} + \hspace{1pt} R_{\mu \nu} \hspace{1pt}
\big(\hspace{1pt} i\,\dot \xi^{\mu} \xi^{\nu}
+ \lambda_2^{\mu}  \lambda_2^{\nu}\hspace{1pt}\big)
+ R^\prime_{i j} \hspace{1pt} 
\big(\hspace{1pt}i\,\dot \xi^{i} \xi^{j} 
+ \lambda_2^{i}  \lambda_2^{j} \hspace{1pt}\big)\Big]
+ \frac 12 R^\prime_{i j} \hspace{1pt}
\psi_0^{i} \psi_0^{j} \;. \nn
\eea
The path-integral yields then
\bea
Z \a=\a -\frac 18 \int \! d x_0^\mu \! \int \! d\psi_0^{\mu} \,
\frac {\det\nolimits_P(i \eta_{\mu \nu} \partial_\tau)
\det\nolimits_A(i \eta_{\mu \nu} \partial_\tau + R_{\mu \nu})}
{\det\nolimits_P(\eta_{\mu \nu} \partial_\tau^2
+ i R_{\mu \nu} \partial_\tau)} \nn \\ \a\;\a \hspace{20pt}
\frac {\det\nolimits_A(i \eta_{i j} 
\partial_\tau) \det\nolimits_P(i \eta_{i j} \partial_\tau + R^\prime_{i j})}
{\det\nolimits_A(\eta_{i j} \partial_\tau^2 
+ i R^\prime_{i j} \partial_\tau)}\,
\int \! d\psi_0^{i} \, \exp \left\{\frac t2 R^\prime_{i j} 
\psi_0^{i} \psi_0^{j}\right\} \;. \nn
\eea
Finally, evaluating the determinants one recovers (\ref{t}) \cite{ss1}.

\section{Anomalous couplings}

Using the results (\ref{s}) and (\ref{t}), the quantum anomalies on 
$n$ parallel Dp-branes and/or Op-planes are found to be
\bea
\a\a I_{BB} = \ch_{\bf n \otimes \bar n}(F) \wedge 
\frac{\widehat{A}(R)}{\widehat{A}(R^{\prime})} 
\wedge e(R^{\prime}) \;, \nn \\
\a\a I_{BO} = \ch_{\bf n \oplus \bar n}(2 F) \wedge 
\frac{\widehat{A}(R)}{\widehat{A}(R^{\prime})} 
\wedge e(R^{\prime}) \;, \nn \\
\a\a I_{OO} = - \hspace{1pt}\frac 18 \hspace{2pt}
\frac{\widehat{L}(R)}{\widehat{L}(R^{\prime})} 
\wedge e(R^{\prime}) \;.
\label{anof}
\eea
On the other hand, assigning the anomalous couplings
\be
S_{B,O} = \sqrt{2 \pi} \int C \wedge Y_{B,O} \;,
\label{ano}
\ee
one gets, according to (\ref{inf}), the classical inflows
\bea
\a\a I_{BB} = - Y_B \wedge \bar Y_B \wedge e(R^{\prime}) \;, \nn \\
\a\a I_{BO} = - \big(Y_B \wedge \bar Y_O + Y_O \wedge \bar Y_B \big)
\wedge e(R^{\prime}) \;, \nn \\
\a\a I_{OO} = - Y_O \wedge \bar Y_O \wedge e(R^{\prime}) \;.
\label{inff}
\eea
Thanks to the property 
$\sqrt{\widehat{A}(R)} \sqrt{\widehat{L}(R/4)} = \widehat{A}(R/2)$,
the relevant (D+2)-form component of (\ref{anof} and
(\ref{inff}) are compatible, and anomaly cancellation requires
\be
Y_B = \ch_{\bf n} (F) \wedge 
\sqrt{\frac{\widehat{A}(R)}{\widehat{A}(R^{\prime})}} \;,\;\;
Y_O = - 2^{p-4} \sqrt{\frac{\widehat{L}(R/4)}
{\widehat{L}(R^{\prime}/4)}} \;.
\label{y}
\ee
Notice that for non-trivial embeddings
and multiple branes, the pulled-back curvatures 
depend also on the gauge connection 
(see for instance \cite{myers}).

The presence of anomalous couplings of the form (\ref{ano}) for D-branes 
and O-planes has been also predicted in particular cases using string 
dualities \cite{bsv,djm}.
Their actual occurrence in the form (\ref{y}) has been demonstrated 
in \cite{mss} through a direct string theory computation, by factorizing
RR magnetic interactions between D-branes and O-planes, encoded in 
one-loop amplitudes on the annulus, M\"obius strip and Klein bottle
surfaces. These couplings have been also checked though tree-level 
computations on the disk and the crosscap \cite{cr,s}.

\section{String theory computation}

Interestingly, the anomaly inflow mechanism on D-branes and O-planes 
can be analyzed directly in string theory, where tadpole cancellation 
guarantees overall finiteness and implies anomaly cancellation. 
Recall that one can compute anomalies by evaluating 
amplitudes with external photons and/or gravitons, one 
of them being pure gauge. This measures the clash of gauge invariance and 
gives directly the anomaly. Only the CP-odd part of potentially divergent 
diagrams can contribute. In string theory, these are the annulus, the 
M\"obius strip and Klein bottle amplitudes in the RR odd spin-structure. 

The amplitudes we want to compute have the form
$$
{\cal A} = \int_0^\infty \! dt \, 
\big\langle\,V^{phy.}_1\,V^{phy.}_2\, ... \, V^{phy.}_n\,
V^{unphy.}\,(T_F + \tilde{T}_F)\,\big\rangle \;.
$$
The insertion of $T_F + \tilde{T}_F$ is due to the gravitino zero mode, 
and the vertices must have total superghost charge $-1$. Take all the 
$V^{phy.}$'s in the $0$-picture, with an arbitrary transverse polarisation 
$\xi_M$ or $\xi_{MN}$, and $V^{unphy.}$ in the $-1$-picture, with 
a longitudinal polarisation given by  $\xi_M = p_M \eta$ or 
$\xi_{MN} = p_M \eta_N + p_N \eta_M$.
Interesting enough, the latter can then be written as a supersymmetry 
variation, $V^{unphy.} = \big[Q + \tilde Q, \hat V^{unphy.}\big]$.
Using standard arguments, one can then move $Q + \tilde Q$ onto the other 
operators in the correlation. One gets no effect on the $V^{phy.}$'s, since
they are supersymmetric, but the supercurrent is changed to the 
energy-momentum tensor,  
$\big[Q + \tilde Q,T_F+\tilde{T}_F\big] = T_B+\tilde{T}_B$.
The net effect of $T_B+\tilde{T}_B$ is to take the derivative 
of the remaining correlation with respect to the modulus $t$, and
one is then left with a total derivative in moduli space:
\be
{\cal A} = \int_0^\infty \! dt \, \frac d{dt} \,
\big\langle\,V^{phy.}_1\,V^{phy.}_2\, ... \,V^{phy.}_n\,
\hat V^{unphy.}\,\big\rangle \;.
\label{anostr}
\ee

In consistent models, this total anomaly has to vanish, reflecting a 
cancellation between one-loop anomalies and tree-level inflows associated 
to the same surface. 
At finite $p$'s, only the ultraviolet boundary $t \rightarrow 0$ can 
contribute and has to vanish by itself. The computation is still difficult, 
but fortunately, to get a field theory interpretation, it is enough to 
restrict to the leading order in $p \rightarrow 0$. 
In this limit, the correlation becomes $t$-independent and yields at the 
same time the anomaly and the inflow. Moreover, since the correlation 
vanishes unless all the fermionic zero modes are inserted, one can 
use \cite{mss,ss1}
\bea
V^{eff.}_\gamma \a=\a \oint\!d\tau\,F \;, \nn \\
V^{eff.}_g \a=\a \oint\! d^2z\,R_{MN} 
\left[X^M (\partial + \bar \partial) X^N \! + 
(\psi - \tilde \psi)^M (\psi - \tilde \psi)^N \right] \;.
\label{vert}
\eea
This holds both for physical and unphysical vertices, with
\bea
\a\a F^{phys.} = \frac 12 F_{\mu \nu} \, \psi_0^\mu \psi_0^\nu \;,\;\;
R^{unphys.}_{MN} = \frac 12 R_{MN\mu\nu} \, \psi_0^\mu \psi_0^\nu \;, \nn \\
\a\a F^{unphys} = \eta \;,\;\; R^{unphys.}_{MN} = p_M \eta_N + p_N \eta_M \nn
\;.
\eea

The generating functional of (\ref{anostr}) is a partition function 
twisted by the interactions (\ref{vert}) in the backgrounds $F+\eta$ and 
$R_{MN} + p_M \eta_N + p_N \eta_M$. The correct number of physical vertices 
is automatically selected, the unphysical one being obtained by 
restricting to the term linear in $\eta$. Not too surprisingly, the only 
role of the unphysical vertex is to take the descent of the remaining 
partition function, and the anomaly polynomial is given by $I = Z^\prime$
\cite{ss2}.

It is straightforward to applying this general result to standard D-branes 
and O-planes. One finds ($\Omega_I = \Omega \, I$ is the $T$-dual of the 
world-sheet parity $\Omega$)
\bea
\a\a I_{BB} = Z_A^\prime \hspace{3pt} 
= \frac 14 \,\Tr^\prime_{R}\, \big[\hspace{1pt}(-1)^F\, 
e^{\raisebox{1pt}{$\scriptstyle{- t H}$}}\hspace{1pt}\big] \;, \nn \\
\a\a I_{BO} = Z_M^\prime = \frac 14 \,\Tr^\prime_{R}\,
\big[\hspace{1pt}\Omega_I\,(-1)^F\, e^{\raisebox{1pt}{$\scriptstyle{- t H}$}}
\hspace{1pt}\big] \;, \nn \\
\a\a I_{OO} = Z_K^\prime \hspace{2pt} = \frac 18 \,\Tr^\prime_{RR}\,
\big[\hspace{1pt}\Omega_I\,(-1)^{F+\tilde F}\,
e^{\raisebox{1pt}{$\scriptstyle{- t H}$}}
\hspace{1pt}\big] \;. \nn
\eea
These are supersymmetric indices, and only massless modes do contribute. 
Effectively, one recovers precisely the SQM models seen before, reproducing 
therefore the same results for the anomalies and the anomalous couplings.

One can apply this general approach also in more complicated cases, 
like for instance D-branes, O-planes and fixed-points in orientifold 
models \cite{ss2}. This provides an efficient tool to analyse in detail 
the complicated G-S mechanism of anomaly cancellation in this kind of
models.

\vspace{5mm}
\par \noindent {\Large \bf Acknowledgments}
\vspace{3mm}

\noindent
Work supported by EEC under TMR contract ERBFMRX-CT96-0045 and 
by the Nederlandse Organisatie voor Wetenschappelijk Onderzoek (NWO).

\end{document}